\documentstyle{laa}
\def\gapprox{\;\rlap{\lower 2.5pt            
 \hbox{$\sim$}}\raise 1.5pt\hbox{$>$}\;}       
\def\lapprox{\;\rlap{\lower 2.5pt            
 \hbox{$\sim$}}\raise 1.5pt\hbox{$<$}\;}

\def\lsim{\, \lower2truept\hbox{${< \atop\hbox{\raise4truept\hbox{$\sim$}}}$}\,}
\def\gsim{\, \lower2truept\hbox{${> \atop\hbox{\raise4truept\hbox{$\sim$}}}$}\,}
\begin{document}
\thesaurus{11(11.07.1; 11.19.5); 13(13.09.1; 13.21.1); 09(09.04.1)}
\title {The Observation of the Nearby Universe in UV and in FIR: an evidence 
 for a moderate extinction in present day star forming galaxies}

\author { V. Buat\inst{1,2}, D. Burgarella\inst{1} }
\offprints {V. Buat} 
\institute { Laboratoire d'Astronomie Spatiale du CNRS, BP 8, 13376 
Marseille Cedex 12, France
\and Laboratoire des int\'eractions photons-mati\`ere, Facult\'e des Sciences 
 de Saint J\'er\^ome, 13397 Marseille Cedex 13, France}
\date{ Received; accepted .... }
\maketitle
\begin{abstract} 

We study the FIR and UV-visible properties of star forming galaxies in the
nearby Universe. This comparison is performed using the local luminosity
functions at UV and FIR wavelengths and on individual starburst galaxies for
which photometric data from UV to NIR and FIR are available. 

The FIR and UV luminosity functions have quite different shapes~: the UV 
function exhibits a strong increase for low luminosity galaxies whereas the FIR
tail towards ultra luminous galaxies ($\rm L > 10^{11}  L\odot$) is not detected
in UV. The comparison of the FIR and UV local luminosity densities argues for a
rather moderate extinction in nearby disk galaxies. The galaxies selected to be
detected in FIR and UV are found to be located in the medium range of both
luminosity functions.  

An emphasis is made on starburst galaxies. For a sample of 22 of these 
objects, it is found that the UV (912-3650 $\rm \AA$), the visible (3600-12500
$\rm \AA$) and the NIR (12500-22000 $\rm \AA$) wavelength range contribute
$\sim 30\%$,  $\sim 50\%$ and  $\sim 20\%$ respectively to the total emerging
stellar emission (for a subsample of 12 galaxies for the NIR and  visible
light). The mean ratio of the dust to bolometric luminosity  of these galaxies
is 0.37$\pm$0.22 similar to the ratio found for normal spiral galaxies. Only 4
out of the 22 galaxies exhibit a very large extinction with more than 60$\%$ of
their energy emitted in the FIR-submm range.  The mean extinction at 2000$\rm
\AA$ is found to be $\sim 1.2$ mag although with a large dispersion. The UV,
visible and NIR emissions of our sample galaxies are  consistent with a burst
lasting over $\sim 1$ Gyr. The conversion factor of the stellar emission into
dust emission is found to correlate  with the luminosity of the galaxies,
brighter galaxies having a higher conversion factor. Since our sample appears
to be representative of the mean properties of the galaxy  population in FIR
and UV, a very large conversion of the stellar light into dust emission can no
longer be assumed as a general property of starburst galaxies at least in the
local Universe.  Instead a larger amount of energy emerging from the present
starburst galaxies seems to come from the stars rather than from the dust. 

We compare the UV properties of our local starburst galaxies to those  of
recently detected high redshift galaxies. The larger extinction found in the
distant galaxies is consistent with the trend we find for the nearby starburst
galaxies namely the brighter the galaxies the lower the escape fraction of 
 stellar light.
                  
\end{abstract}

\section{Introduction}

One of the most important challenge of modern astronomy is the detection of
young primeval galaxies. Indeed, very significant progress has been made  with
the detection of very high redshift galaxies either from ground based 
observations (Steidel et al. 1996b) or in the Hubble Deep Field (e.g. Steidel 
et al. 1996a, Lowenthal et al. 1997).  

In order to understand the properties of high redshift galaxies and study the
cosmological evolution of star-forming galaxies, it is crucial to properly
characterize the properties of starburst galaxies in the local Universe.
These nearby galaxies are forming stars with a very high rate and it it
actually important to analyse their emission over the entire spectral range
(from UV to FIR-submm) to study the efficiency of the dust extinction and know 
what is the spectral range (UV, visible, NIR or FIR) where most of their 
energy is emitted. If high redshift star forming galaxies are similar to their 
low-z counterparts studying the latter will bring some clues to 
detect the former. We can wonder whether the observation of the rest-frame UV 
continuum is the best way to detect high redshift galaxies or if the high
obscuration from dust makes them emit more energy in the FIR.

IRAS discovered infrared bright galaxies with prodigious star formation rates, 
a very high extinction and therefore a low optical flux (Sanders \& Mirabel  
1996).  The most luminous FIR galaxies are produced by strong interactions or 
merging of molecular gas-rich galaxies which induce enormous starbursts. Such 
objects might be the progenitors of elliptical galaxies (Kormendy \& Sanders 
1992). In a "bottom-up" scenario of galaxy formation, numerous starbursts 
induced by merging are expected and the bulk of their emission would be in the 
FIR redshifted in the submm (e.g. van der Werf \&  Israel 1996). Mazzei et al.
(1994) predict that more than 90 \% of the energy emitted by a starburst is in
the FIR range during the first Gyr. This percentage rapidly drops to reach
$\sim 30\%$ for $\sim 5$ Gyr- old objects. Models aimed at explaining the
galaxy counts in optical and FIR predict that during intense phases of star
formation the quasi totality of the stellar light is absorbed and re-radiated
in the FIR wavelength range (Franceschini et al. 1994, Pearson \& 
Rowan-Robinson 1996).

Conversely, considerable effort has been carried out in the UV-optical study 
of star forming galaxies for some years. Calzetti, Kinney and co-workers have
extensively used the IUE spectra of star-forming galaxies complemented with
optical and IR data to characterize the star formation history and the
extinction occurring in the central regions of these objects (Calzetti et al.
1994, Calzetti et al. 1995, Calzetti 1997a). Meurer et al. (1995) have used
Faint Object Camera on board the {\it Hubble Space Telescope} HST-FOC
observations to study the morphology of some starburst galaxies. From
these studies, a foreground distribution of dust and a rather grey extinction 
curve seems to be able to explain the spectral distribution of the central 
regions of starburst galaxies. The extinction found by Meurer et al. (1995) 
for nearby starburst galaxies is rather low: at 2000 $\rm \AA$ it lies between 
0.08 and 1.9 mag (excluding NGC7552 at 3.13 mag). Nevertheless these studies 
deal with the central parts of starburst galaxies and may well not be valid for 
 the global emission of these objects at least when longer wavelengths than UV
are concerned (Buat et al. 1997). 

Analyses of the UV-optical and FIR global emissions of nearby spiral and 
irregular galaxies selected to be observed both in UV and in FIR led to a 
rather low extinction (Xu \& Buat 1995, Buat \& Xu 1996, Wang \& Heckman 1996). 
An important result of these studies is that the UV non-ionizing stellar 
emission is likely to be the major cause of dust heating. The contribution  of
OB stars to the dust heating is estimated to amount to about 20$\%$ of the
total FIR emission in the Milky Way (Cox \& Mezger 1989), almost the same
contribution of the ionizing  radiation to the dust heating is found for spiral
galaxies (Xu \& Buat, 1995)  and for  starburst objects (Calzetti et al. 1995).  
The comparison  between the FIR emission (dust re-radiation) and the UV and
optical emission  (escaped stellar light) constrains the extinction. As a
consequence the FIR to  UV continuum ratio is a powerful indicator of the
extinction occurring in  galaxies (Meurer et al. 1995, Buat \& Xu 1996, Wang \&
Heckman 1996).

Recently, HST imaging of very high redshift galaxies complemented when 
possible by spectroscopic observations with the Keck Telescope have led to the
discovery at high redshift (z$\sim 2-3$) of compact star forming galaxies with
a moderate size and a strong rest-frame UV emission (Steidel et al. 1996a) with 
sometimes more diffuse extended structures  Lowenthal et al. 1997). Depending
on the intrinsic UV spectrum adopted, the  average extinction estimated at 1600
$\rm \AA$ from the rest-frame UV spectral  energy distribution of these
galaxies is of the order of 1.7 to 3 mag (Meurer  et al. 1997, Calzetti 1997b).
These significant average extinctions are  therefore larger than the values
estimated for nearby starburst galaxies.  However, it must be noted that these
high redshift galaxies are very luminous  ($\rm M_B < -21$) when compared to
the mean luminosity of nearby starburst  galaxies studied by Meurer et al.
(1995) ($\rm <M_B> = -18.6$) and the  extinction is known to correlate with the
luminosity of galaxies  (Giovanelli et al. 1995, Wang \& Heckman 1997). 
Moreover, the selection biases are very strong towards very luminous galaxies
with a strong UV continuum and it cannot be excluded that high redshift
galaxies almost entirely hidden by the dust are missing from these observations
in the rest-frame UV (Mobasher et al. 1996, Burigana et al. 1997). 

The selection biases in the recent detections of high z galaxies are difficult 
or even impossible to quantify in the absence of similar observations at other
wavelengths corresponding to longer than UV rest frame emissions (NIR or FIR). 
A first step is to estimate the importance of such a bias in the
local Universe. Such a study is also crucial to compare the properties of  
high redshift galaxies to those in the nearby Universe. 
 
 At this aim  we will adopt a global approach to study the local Universe which 
consists in comparing the luminosity functions and the luminosity densities in 
UV and FIR.  The UV wavelength range is particularly interesting since it is  a
tracer of the recent star formation rate as already mentioned and  observations
in the visible  range of high z ($>2$) galaxies correspond to their UV rest frame
emission. 
 More specifically we will compare the amount of energy locked up in FIR to 
 the amount
of energy directly emitted in UV in the local Universe. The comparison of these 
global values with  individual galaxies selected to be observed both in UV 
and FIR will allow to discuss how such samples of individual galaxies are 
representative of the mean properties of the local Universe. 

We will also investigate the specific case of a sub sample of nearby starburst
 galaxies detected in UV, visible, NIR and FIR in order to compare their 
global dust and stellar emission and to estimate what fraction of the emission
of stars is converted into dust emission as well as the relative contribution 
of the UV, visible and NIR spectral ranges to the observed stellar emission. 
Such estimates will lead to predict what spectral range is more favorable for
the detection of high redshift starbursts under the hypothesis that they are
similar to their nearby counterparts. The main limitation to this approach 
is that we deal with global fluxes integrated over the galaxies whereas the
starburst often occurs in the central parts. Moreover the galaxies at high
redshift so far detected seem to have compact morphologies. Nevertheless as 
 it will be shown in section 4  a  large fraction of the 
 UV emission of a starburst galaxy  is likely to come from the starburst itself making 
valid a study on the global fluxes as soon as this wavelength is concerned. 
It is also the case for the FIR emission since the UV (ionizing and 
 non ionizing) emissions is the major contributor to
the dust heating), especially in starbursting objects. Obviously, more care
must be taken when dealing with  the visible and NIR emission: at these 
 wavelengths  the contribution of the underlying old stellar population present 
 in local starburst galaxies is very large even dominant. Endly, available FIR
fluxes on large samples are integrated over the galaxies due to the poor 
 resolution of the IRAS  satellite and dealing with global fluxes allows a
reliable comparison of the emission of galaxies at different wavelengths.

Beyond the detection of high redshift star forming galaxies it is necessary to
estimate a quantitative star  formation rate (SFR) for these galaxies. The
deduction of such a quantity from  the observed rest-frame UV continuum relies
almost entirely on the amount of  the extinction with only a moderate 
dependence on the star formation history  (e.g. Meurer et al. 1997, Calzetti
1997). More specifically  after $\sim 5~10^7$ years of constant star  formation
rate the UV flux (912-3650 $\rm \AA$) reaches 80 $\%$ of its stationnary value 
calculated for $\rm 10^{10}$ years of constant star formation (from  
Bruzual \& Charlot  1993). From a  global energetic budget, we will try to
constrain the amount of extinction and bring some clues to this difficult 
problem.

\section{The data on individual galaxies}

Our study is based on samples of galaxies detected in UV and for which FIR data 
are available. We have used the UV observations ($\rm \sim 2000  \AA$) of the
balloon borne telescopes SCAP and FOCA (Donas et al. 1987, 1990, 1995) together
with those obtained with FAUST (Deharveng et al. 1994). The FIR data come from
the IRAS database.

First, we will re-consider a first sample of 152 galaxies already compiled
(Buat \& Xu 1996) in order to investigate the selection biases inherent to our
selection. 

Second, we built a sample of starburst galaxies.  These starburst galaxies 
are selected according to their far-infrared color: $\rm f_{60}/f_{100} >
0.55$. It ensures a dust temperature larger than 36K
assuming a single temperature thermal model and a dust emissivity index equal
to 1 in agreement with the definition of starburst galaxies adopted by Pearson
and Rowan-Robinson (1996). These starburst galaxies are selected to be observed
in UV photometry and by IRAS at  60 and 100   $\rm \mu m$.  For galaxies
belonging to the Coma, Abell 1367 or Virgo clusters,  the visible and NIR data
come from Gavazzi et al. (1996a,b), Gavazzi \& Boselli (1996 and references
therein), Boselli et al. (1997) and Boselli et al. (1998, in preparation); the
FIR data are compiled by Gavazzi (private communication). The data for non
cluster galaxies are from de Vaucouleurs et al. (1991, RC3) and the IRAS 
database. Active galaxies are excluded. 22 galaxies are selected and gathered
in Tab. 1.  Most of the  galaxies of this sample (and all the nearest ones not
belonging to Coma or Abell 1367  clusters) are reported in the literature as 
experiencing a starburst. 

The distances of nearby galaxies are taken from Tully (1988). The distances to 
A1367 and Coma clusters are taken to 87 and 92 Mpc respectively. When the
galaxies  do 
not belong to a cluster and are not listed in  Tully's catalog, 
 their radial velocity are taken from the RC3 and the
distances computed assuming $\rm H_0 = 75 km/s/Mpc$ in consistency with Tully; 
h is defined as $\rm H_0/100$.

\begin{table}
\caption[]{Starburst galaxies ($\rm f_{60}/f_{100}~>~0.55$) detected to be
observed in FIR and UV. (1) Name NGC, IC or Zwicky by order of preference,
(2) morphological type using the RC3 nomenclature, (3) absolute B magnitude,
(4) 60 to 100 $\rm \mu$m flux ratios and (6) ratios of the FIR (40-100 $\rm 
\mu$m) to UV (defined as $\rm \nu F_{\nu}$ at 2000 $\rm \AA$) luminosities }
\begin{flushleft}
\begin{tabular}{llllll}
\hline
Name & Type & $\rm M_b$ & $\rm f_{60}/f_{100}$ & $\rm L_{FIR}/L_{UV}$ & 
NIR data\\
\hline
 NGC3034 &   8  &    -19.52 &  0.64   &   172.72& $~$ \\
 
 NGC3353 &   3  &    -17.92 &  0.78   &    3.98 \\
 NGC3913 &   5  &    -19.01 &  0.59   &    2.72 &  H \\
 NGC4194 &  10  &    -20.00 &  0.98   &    25.53 \\
 NGC4383 &   1  &    -18.50 &  0.66   &    3.85   &  J H K \\
 NGC4424 &   1  &    -18.71 &  0.56   &    8.19   &    H   \\
 NGC4519 &   5  &    -18.79 &  0.58   &    1.48   &     J H K \\
 NGC4532 &  8   &   -18.81  & 0.58    &   2.46    &    J H K  \\
 NGC4670 &  8   &   -17.19  & 0.59    &   1.26    &      H   \\        
 NGC4922 &  8   &   -19.72  & 0.81    &   107.85  &      H   \\
 NGC5253 &  8   &   -16.76  & 1.09    &   2.05    &    $~$\\
 NGC5477 &   9  &    -14.68 &  0.58   &    0.43   &   $~$\\
 NGC7673 &   5  &    -20.20 &  0.64   &    1.65   &  $~$\\       
 NGC7677 &   4  &    -19.46 &  0.67   &    3.60   & $~$\\
 IC732   &   8  &    -18.76 &  0.65   &    243.66 &       H\\
IC3258 &  10  &    -17.51 &  0.56   &    0.65 & $~$ \\
IC3576 &   9  &    -17.12 &  0.60   &    0.47 & $~$ \\ 
Z97068  &   4  &    -20.03 &  0.55   &    7.66   &     J H K\\
 Z97079  &   8  &    -18.83 &  0.60   &    1.47   &     J H K\\
 Z160076 &   5  &    -18.95 &  0.76   &    0.65   &       H  \\
 Z160106 &   8  &    -19.48 &  0.73   &    9.05   &     J H K \\
 Z160139 &   8  &    -18.80 &  0.62   &    1.09   &       H  \\
\hline
\end{tabular}
\end{flushleft}
\end{table}

\section {Comparison of the local luminosity functions at 60 $\mu$m and 
2000 $\rm \AA$}
 
\subsection {The luminosity functions at 60 $\mu$m and 2000 $\rm \AA$} 

The local FIR luminosity function at 60 $\rm \mu m$ has been determined from 
IRAS observations (e.g. Soifer \& Neugebauer 1991, Saunders et al. 1990, 
Koranyi \& Strauss 1997). Saunders et al. used a compilation of samples and 
determined an analytical expression of the luminosity function at 60 $\rm \mu
m$. This analytical form will be used hereafter. 

Recently, the spectroscopic follow-up of ultraviolet observations at 2000 $\rm
\AA$ allowed to estimate a local UV luminosity function (Milliard et al. 1997, 
Treyer et al. 1997). In spite of the small number of galaxies involved in the 
determination of the UV luminosity functions we compare the shapes of these 
functions at 60 $\mu$m and 2000 $\rm \AA$ (Fig.~1). In this figure the UV 
 luminosity function is represented by the Schechter function fitted by Treyer
et al.. The largest observational 
errors quoted by Treyer et al. reach 0.8 in units of $\rm 
\log(\Phi)$ (vertical axis in Fig.~1) at the faint and bright ends of the UV 
luminosity function whereas the error bars reported 
by Saunders et al. correspond to $\sim 1.$ in units of $\rm \log(\Phi)$ at the 
faint end and $\sim 0.5$ at the bright end of the $\rm 60\mu m$ luminosity 
function. Fig.~1 clearly shows that the two functions exhibit very different 
shapes at a high level of confidence~: the UV luminosity function rises sharply 
for low luminosity galaxies while the tail of FIR bright galaxies has no 
counterparts in UV. Note however that Milliard et al. (1997) do not observe the
 knee in the UV luminosity function with a rather monotonic decrease towards
large luminosity galaxies.
A more thorough comparison of the luminosity functions is
beyond the scope of this paper and devoted to a subsequent work. 

\begin{figure}
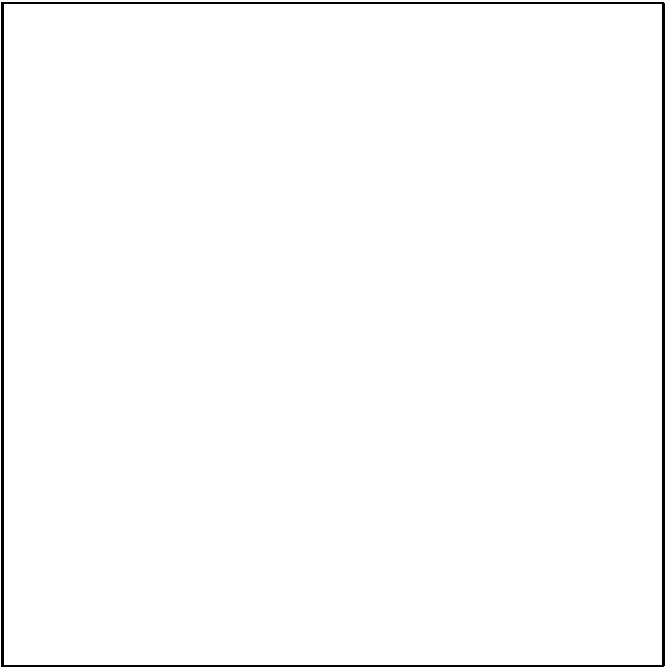

\picplace{8.8truecm}
\caption[]{ Luminosity functions at $\rm 60\mu m$ (solid line) and   2000 $\rm
\AA$ (dotted line) with h=0.75.  The luminosity of the galaxies at the wavelength
considered ($\rm 60\mu m$ or  2000 $\rm \AA$) is reported on the x-axis. The
luminosities expressed in solar units are defined  as $\rm \nu L_{\nu}$}
\end{figure}

The study of disk galaxies detected in UV and FIR led to a moderate extinction
even in UV around 0.5-1 mag (Buat \& Xu 1996, Wang \& Heckman 1996, Buat 1997) . However, the impact of these studies is 
limited by the difficulty to estimate and understand the selection biases 
inherent to the fact that the galaxies must be detected in UV and FIR. We can 
test these potential biases by comparing the luminosity distributions of 
samples of galaxies selected to be detected in UV and FIR to more global 
luminosity functions. At this aim, we use the largest sample compiled by us
which consists in 152 spiral and irregular galaxies (Buat \& Xu 1996). In 
Fig.~2 are reported the luminosity distributions of these galaxies at 60 
$\mu$m and 2000 $\rm \AA$. By a comparison with the luminosity functions of
Fig.~1, we can see that we select galaxies near the "knee" of the luminosity
functions. The sharp decrease at low luminosities is due to the detection
limits. The luminosity distribution of our sample of starburst galaxies
presented in Tab.~1 is also reported in Fig.~2. The distribution of their UV
luminosity is consistent with that of the former sample. These galaxies appear
to have in average a slightly larger luminosity at 60 $\mu$m in agreement with
the result of Saunders et al. (1990). In conclusion the galaxies we study have
medium luminosities at each wavelength. As we may have expected, the effects
due to the UV-based selection of galaxies are counterbalanced by the necessity
to have these objects also detected in FIR. We can therefore conclude that
{\it galaxies selected in such a way can be considered as representative of the
mean population of galaxies}.

\begin{figure}
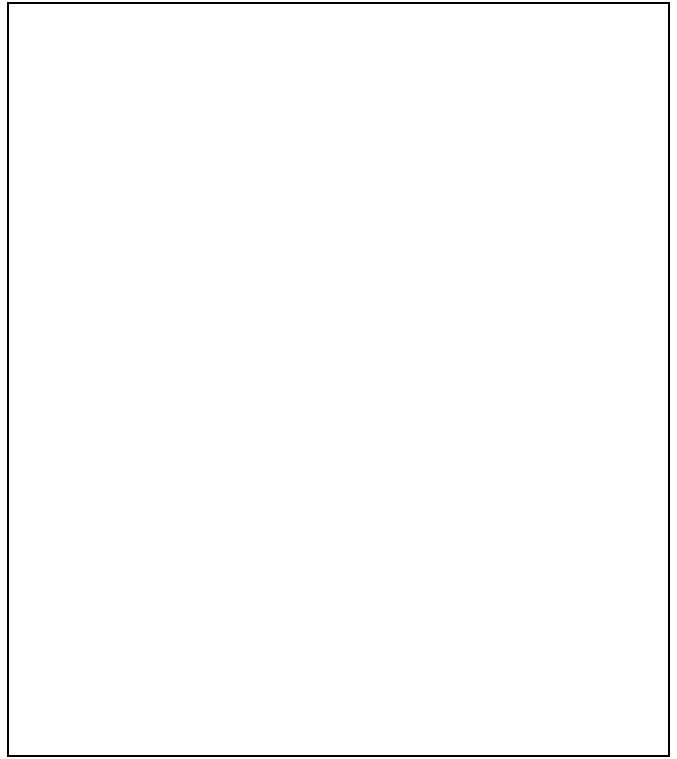

\picplace{10truecm}                       
\caption[]{Luminosity distributions at (a) $\rm 60\mu m$ and (b) and 2000 $\rm 
\AA$ for galaxies detected at both wavelengths. The distributions of starburst
galaxies are plotted with a dotted line}
\end{figure}

\subsection {The UV and FIR local luminosity densities}

It is important to know what fraction of the energy emitted by a star forming
galaxy will fall in the UV or FIR range. This result will allow us 
to predict which of the two spectral domains is best suited to search for 
high-z star-forming objects. Moreover, the comparison of the FIR and UV
emissions of galaxies has been recognized to be a powerful way to estimate the
extinction (e.g. Meurer et al. 1995, Buat \& Xu 1996, Wang \& Heckman 1996),
the UV non-ionizing flux being found to dominate the dust heating (Buat \& Xu
1996 but see Trewhella et al. 1997). In the following, we will estimate this
ratio deduced from local luminosity densities and for individual galaxies.

Saunders et al. (1990) estimate the local luminosity density at 60 $\rm \mu m$ 
(defined as the product of the frequency $\rm \nu$ by the luminosity density 
per unit frequency at 60 $\rm \mu m$) and in the FIR (40-120$\rm \mu m$)
range: 

$$\rm \rho_{60} =  (4.2\pm 0.4)\times 10^7~ h~ L\odot/Mpc^3$$  
$$ \rm \rho_{FIR} =  (5.6\pm 0.6)\times 10^7~ h~ L\odot/Mpc^3$$

The local luminosity density at 2000 $\rm \AA$, defined as the product of the
frequency $\rm \nu$ by the luminosity density per unit frequency at 2000 $\rm
\AA$, is found to be (Milliard et al. 1997, Treyer et al. 1997)~:

$$ \rm \rho_{UV} = (6.3\pm 2.7) \times 10^7~ h~ L\odot/Mpc^3$$  

From the values quoted above we deduce~:

$$\rm \rho_{FIR}/\rho_{UV} = 0.9\pm 0.5$$

The global ratio of the luminosity densities can be compared to the ratio
between the FIR and UV emissions of individual galaxies. First, we use the
sample of 152 spiral and irregular galaxies (Buat \& Xu 1996) for which we
find~:

$$\rm <L_{FIR}/ L_{UV}> = 3 $$ 

Given the very large dispersion ($\sigma$ = 5) of the ratio we have also
measured the median of the distribution which is found equal to 1.6. Second,
the case of starburst galaxies is also explored. Tab.~1 lists the ratio between
the FIR and UV  luminosities for these galaxies. These values are also very
dispersed extending from 0.4 to 244 with a median value equal to 2.5. Therefore
from the  comparison of the FIR to UV ratio the selected  starburst galaxies 
seem to be slightly more obscured than the mean of spiral and irregular
galaxies as already found from the luminosity distributions. Nevertheless such
a conclusion must be taken with caution  since the calibration of  the FIR to
UV ratio in terms of extinction might well depend on galaxy properties like the
dust temperature or the star formation history. 

\begin{figure}
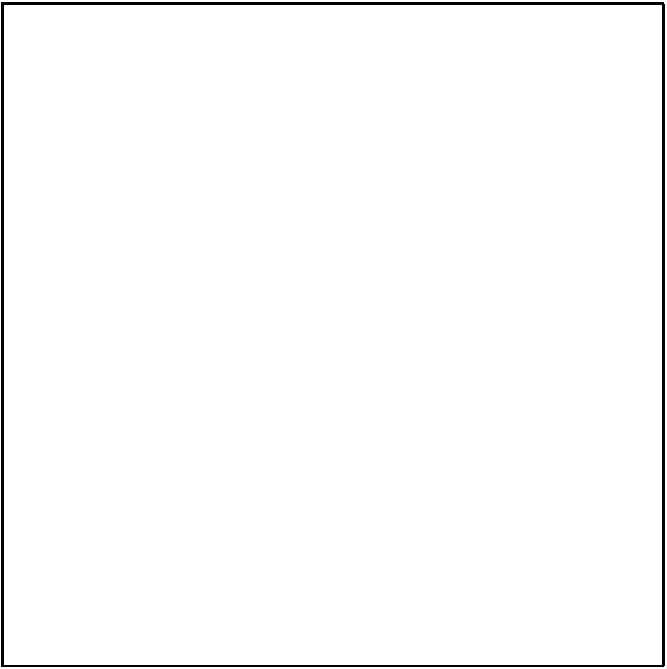

\picplace{8.8truecm}
\caption[]{$\rm L_{UV}/L_{60}$ luminosity ratio as a function of the
$\rm L_{60}$ luminosity at $\rm 60\mu m$}
\end{figure}

An other way to compare the FIR and UV emissions of individual galaxies to the
luminosity densities consists in estimating the local UV density using the
luminosity function at 60$\rm \mu m$ and the $\rm L_{UV}/L_{60}$ ratio found
for individual galaxies. Such an approach has the advantage to be independent 
 of the shape of the UV luminosity function which is not yet very secure 
  due to the small number of galaxies involved (Milliard, 
 private communication). 
 A correlation is found between the $\rm L_{UV}/L_{60}$
ratio and the $\rm L_{60}$ luminosity as shown in Fig.~3. A linear regression 
gives~:

$$\rm  \log(L_{UV}/L_{60}) = 18.56 - 0.44~ \log(L_{60})$$ 

\noindent where the independent variable $\rm L_{60}$ is expressed in 
$\rm erg/s$. The local luminosity density at 2000 $\rm \AA$ can be calculated 
as~:

$$\rm \rho_{UV} = {\int_{L_{60~min}}^{\infty}} \phi(L_{60})\cdot 
(\rho_{UV}/\rho_{60})\cdot d(\log(L_{60}))$$ 

\noindent where $\rm \phi(L_{60})$ is the present epoch $\rm 60 \mu m$ 
luminosity function estimated by Saunders et al. (1990) and $\rm
\rho_{UV}/\rho_{60}$ is equal to $\rm L_{UV}/L_{60}$. 

The lower boundary of the integral $\rm L_{60~min}$ has to be estimated
according to the limiting absolute UV magnitude used to evaluate the local
luminosity density. Treyer et al. (1997) adopt $\rm M_{UV}<-16$  for $\rm H_0 =
100 km/s/Mpc$ which translates to $\rm M_{UV}<-16.6$ for  $\rm H_0 = 75
km/s/Mpc$. Using the regression formula between   $\rm\log(L_{UV}/L_{60})$ and
$\rm log(L_{60})$ we find $\rm L_{60~min} = 1.1~ 10^{41}  erg/s$. Milliard et
al. (1997) have calculated the local luminosity density at 2000 $\rm \AA$ 
with an extrapolation at fainter magnitudes. The difference between the two
estimates of the local luminosity density is within the error bars  estimated
by Treyer et al. (1997). Therefore we will use the lower limit calculated
above to calculate the local UV luminosity density.

We obtain~:

$$\rm \rho_{UV} = 1.3 \times ~10^{7}~ L\odot/Mpc^3$$  

\noindent with $\rm h=0.75$. This value is lower than the observed one: $ \rm
\rho_{UV} = 3.5 \times 10^7 L\odot/Mpc^3$ in agreement with the fact that the
selected galaxies have a larger FIR to UV ratio than that found from the local
luminosity densities. However the dispersion in Fig.~3 is large; we can
estimate the uncertainty in our calculation by considering the regression
taking $\rm \log(L_{UV}/L_{60})$ as the independent variable, the corresponding
regression line is plotted in Fig.~3 (dotted line). This leads to $\rm
\rho_{UV} = 3.5 \times ~10^{7}~ L\odot/Mpc^3$ similar to the observed value 
but the adopted regression clearly overestimates the UV flux of low luminosity
galaxies in the sample.

Therefore, in spite of the uncertainties of our calculation the calculated UV
luminosity density is lower than the observed one  as expected since our
selection lacks the low luminosity galaxies detected in UV but not in FIR
(Fig.~1) and which largely contribute to the local UV density. Once again this
analysis leaves no much room for a lot of galaxies with a very high extinction 
missed by the UV observations. 

From the analysis of the local luminosity densities, a similar amount of energy
seems to be present in the UV and the FIR spectral range. We must note,
however, that this estimate is very crude since the energies are defined as the
product of a monochromatic flux per unit frequency by the observed frequency
independently of the spectral distribution. For individual galaxies a larger
amount of energy is found in FIR as shown in Fig.~3. We will re-discuss this
result in the next section with a more careful analysis  of the spectral energy
distributions of individual starburst galaxies.

\section{The stellar and dust emissions of starburst galaxies}

We now focus on the case of starburst galaxies (Tab.~1). We have seen in the
previous section that our sample can be considered as typical as far as the
luminosity at both wavelengths is concerned.

A rather straightforward way to estimate the extinction in a bolometric sense
(without a wavelength dependence) is to compare the total (stars+dust)  emission
of a galaxy to the dust emission alone. Models for dust extinction including
geometry and radiation transfer are needed to estimate the extinction at each
wavelength but not for such a global approach. Therefore we will not have to
discuss the various sources of dust heating and the relative contribution to
the dust emission. 

At this aim we have to estimate the dust and stellar emissions emerging from 
the galaxies. The comparison of the FIR and stellar properties of galaxies has
been intensively made in the literature (e.g. Disney et al. 1989, Trewhella et
al. 1997) but without accounting for the emission in the UV spectral range.
Actually, the UV non ionizing stellar light from 912 to 3650 $\rm \AA$ is found 
to dominate the dust heating in most galactic disks (Buat \& Xu 1996). 
Moreover, the contribution of the UV flux to the total stellar light produced
in a galaxy can be estimated using synthetic spectra of galaxies. From
population synthesis models (Bruzual \& Charlot 1993) assuming a constant star
formation rate over 1 Gyr, 2 Gyr and 20 Gyr and no extinction, the ratio of the
UV (912-3650 $\rm \AA$) flux to the flux emitted from 912 $\rm \AA$ to 1 $\rm
\mu m$ is equal respectively to 0.66, 0.63 and 0.50. Therefore, an accurate
estimate of the flux emitted by a galactic disk in UV is crucial for an
analysis of the energy budget. 

\subsection{The total dust emission}
        
Dust emits from a few microns to around 1 mm whereas the combination of IRAS
data at 60 and 100 $\rm \mu m$ is a reliable estimate of the dust emission only
in the 40-120 $\rm \mu m$ wavelength range. Nevertheless, using the total dust
emission calculated by Kwan \& Xie (1992) for 11 galaxies, Xu \& Buat (1995) 
showed that a strong anti-correlation exists between the ratio of the total dust  
emission to the FIR (40-120 $\rm \mu m$) one and the FIR color ratio 
$\rm f_{60}/f_{100}$. Here, we re-calibrate the relation with the new FIR/mm 
dust luminosities reported by Andr\'eani and Franceschini (1996). In Fig.4 is 
reported the ratio of the total dust emission to the FIR one for the 29 
galaxies studied by Andr\'eani and Franceschini (we have excluded one Seyfert 
galaxy). A strong anti-correlation is found and a linear regression gives~:

$$\rm F_{dust}/F_{FIR} = -1.7 (\pm 0.2)\cdot 
 \log (F_{60}/F_{100})+0.91(\pm 0.11)$$ 

\begin{figure}
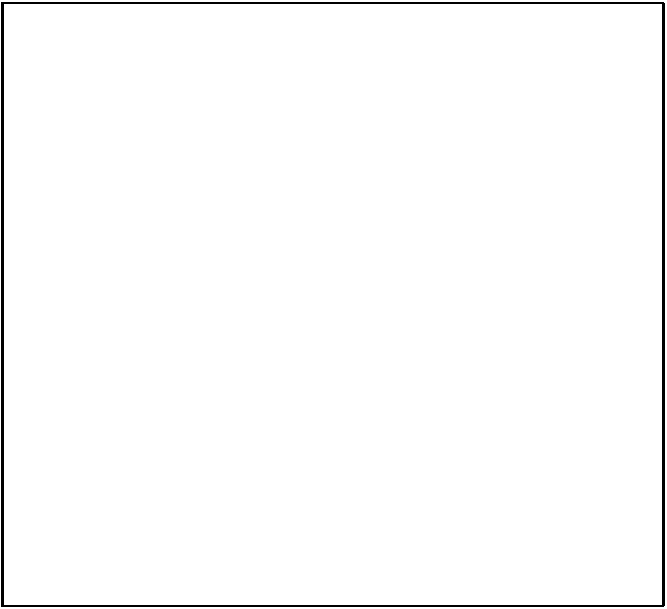

\picplace{8truecm}
\caption[]{The ratio of the total dust to FIR (40 - 120 $\mu$ m) fluxes as a
function of the $\rm F_{60}/f_{100}$ color ratio for the galaxies observed 
by Andr\'eani \& Franceschini (1996)}
\end{figure}

The ratio of the total dust emission and the FIR (40 - 120 $\rm \mu m$) is 
found between 1 and 1.35 for the starburst galaxies of our sample.  This
results differs by only 30$\%$ from that obtained previously   (Xu \& Buat
1995). Andr\'eani and Franceschini use two components (a cirrus and a dust
component)  to fit the fir-submm spectrum whereas Kwan and Xie (1992)
introduced a dust model with a continuous dust temperature distribution.
Modeling the FIR emission with a single temperature component leads to a ratio
between the total dust  flux (1-500 $\rm \mu m$) to the FIR one also 
compatible with our estimate within   the same uncertainty of 30$\%$ whatever
the adopted emissivity law (Helou et al. 1998).  Indeed whereas the
determination of the dust mass is very dependent on the number of dust
components and of their characteristics (temperature, emissivity) the total
dust emission is not  very sensitive to the adopted dust model (e.g. Devereux
\& Young 1992).

\subsection{ The total stellar emission}

The stellar light emerging from a galaxy covers a large spectral range from 
the Lyman limit to a few microns. Therefore the total observed emission of 
stars must be estimated over this spectral range. Since we deal with global
fluxes only photometric data at different wavelengths are available~: UV fluxes
at 2000 $\rm \AA$, U, B, V and in some cases J, H, and K data. These data
constitute broad band spectra which can be used to estimate the global stellar
emission emerging from the galactic disks.

In previous works (Xu \& Buat 1995, Buat \& Xu 1996) we estimated this observed 
stellar emission using empirical spectra indexed on the morphological type of 
the galaxies. Although valid for normal galaxies which are the bulk of the 
objects sampled in these papers, such an approach is not valid for starburst 
galaxies since the UV light distribution in these objects is likely to be 
dominated or at least largely influenced by the starburst and is therefore 
probably not representative of the morphological type of the galaxy.

The method adopted here consists in integrating  the observed broad band 
spectra. However we have to consider all the emission longward the Lyman limit 
with only 7 photometric data in the best cases. The situation is particularly
difficult in the UV range with only one flux at  2000 $\rm \AA$ to cover the
wavelength range 912-3650 $\rm \AA$.  Nevertheless Calzetti et al. (1994) have
shown that the UV spectrum of central regions of starburst galaxies  can be
fitted  by a power law $\rm F_{\lambda} \propto \lambda^{\beta}$  from  912 to
2600 $\rm \AA$; the slope $\beta$ is supposed to depend only on the 
extinction. The extension of this study to the emission of the entire galaxies 
 is not straighforward because of the contribution of the non starbursting
 regions to the total UV emission. A first way to estimate the contribution 
 of the starburst to the integrated UV emission is to use population synthesis
models. Using the models of Bruzual and Charlot (1993) we have added a 
 starburst lasting $10^7$  years and involving a fraction of the total stellar 
 mass to a constant star formation rate over $10^{10}$ years (underlying 
disk population). For a fraction of 10$\%$ of the total
stellar mass created in the burst (Satyapal et al. 1988, Wright et al. 1997) 
 more than 80$\%$ of the flux at 2000 $\rm \AA$ comes from the starburst.  
Another way to estimate the contribution of the starburst is to 
 consider the observations of resolved galaxies. Indeed the central starburst
regions of two galaxies of our
sample (NGC 4670 and NGC 5253) have been observed with the HST at 2200$\AA$
(22"$\times$ 22" field, Meurer et al. 1995). The  integrated 
 UV fluxes over the entire galaxies at $\rm 2000~\AA$ reported in table 1 
 are 1.2 times larger than the flux measured in the HST field: 80 $\%$ of the
 total flux at 2000$\AA$ of these galaxies comes from the central starburst 
region.  
However these galaxies are classified as irregular galaxies and the
contribution of an underlying old population could be expected to be
negligible for such objects. Therefore we have also to envisage 
 the case of earlier type galaxies. To this aim we have considered the
starburst SBbc galaxy NGC 3310 observed both with the Ultraviolet Imaging
Telescope (Smith et al. 1996) and with HST  (Meurer et al. 1995). The central
region observed with HST is found to contain 35$\%$ of the total UV flux 
(Smith et al. 1996). Nevertheless, a large part of 
 the remaining UV emission of NGC3310 
 is 
  located in compact sites of recent star formation (Smith et al. 1996) as it 
 is also 
 the case for other disk galaxies like Messier 33 
  for example (Buat et al. 1994). 
 Therefore except perhaps in the case of very early type galaxies 
 (only 3 galaxies of our sample have a morphological type ealier than Sbc) 
 where the contribution of a population outside the sites of recent star
formation might be quite important, applying the results found for the UV
emission of starbursting regions to the entire galaxies is likely to be a good
approximation.

To 
estimate the slope of the UV spectrum we use the correlation found by 
 Meurer et al. (1995) 
between $\beta$ and the FIR to UV flux ratios. More precisely, we use the curve 
calculated by Meurer et al. (their Fig. 6) using a foreground screen model and
 the 
extinction curve of Calzetti et al. (1994). The unreddened slope of the UV spectrum adopted by Meurer et al. is $\rm \beta
= -2.5$ corresponding to a starburst younger than 100 Myr. We have also tried
the case of a constant star formation rate over 1 Gyr which implies $\rm \beta
= -2$ for the  unreddened UV spectrum and we have checked that the difference 
 in both estimates of the fraction of energy emitted in the UV range reaches 
 only few percents.

Beyond   2600 $\rm \AA$, we directly integrate the spectral energy 
distributions obtained with the U, B, V, J, H and K data and no model fitting.
Only 6 galaxies have been observed at every wavelength. 
From V to K the shape of the SED in log-log 
units is almost linear. 7 galaxies have an H magnitude (without J and K data) 
and for these objects we assume a linear spectral energy distribution from V to 
K. The error due to this approximation are estimated for the 6 galaxies for 
which all the data are available and is found to be 7$\%$ for the estimate of 
the energy contained between the J and K band and only 2$\%$ for the estimate 
of the total stellar flux. The case of the 9 remaining galaxies not observed at 
all in the NIR is more difficult. Nevertheless for the 12 galaxies for which at 
least one NIR band is available a very tight correlation (correlation 
coefficient equal to 0.99) with a slope equal to 0.98$\pm 0.02$ is found 
between the logarithm of the total stellar flux and the logarithm of the flux 
emitted in the visible range from U to V (the latter being chosen as the 
independent variable). The correlation is more dispersed when the UV range is
accounted for. Therefore the total stellar flux is calculated from the flux 
emitted in the (U-V) range by forcing the slope of the linear 
regression to 1. We obtain $\rm \log(F_{star}) = \log (F_{U-V}) + 0.90$. 
The error is estimated to be 12 $\%$ using the 15 galaxies for which NIR data 
are available. 

In Table 2 are reported the stellar and dust fluxes estimated for each starburst 
galaxies together with the contribution of the UV (912-3650$\AA$), visible 
(3650-12500$\AA$) and NIR (12500-22000 $\rm \AA$) wavelength ranges. The 
major contribution to the total stellar flux is from the visible wavelengths 
(from U to J) with $51\pm 11\%$, $32\pm 19 \%$ of the emitted energy comes 
from the UV and $19\pm 11 \%$ from the NIR. Note however that the very large 
contribution of the visible wavelengths is mainly due to the fact that the 
range is extended up to 12500 $\rm \AA$ due to the lack of data between the V 
and J band. 

\begin{table}
\caption[]{Stellar and dust emission of the selected starburst galaxies. The 
fluxes are in $\rm erg/cm^2/s$. $\rm F_{star}$ is the total stellar flux
emerging from the galaxies, $\rm F_{UV}$ is the stellar flux between 912 and
3650 $\rm \AA$, $\rm F_{vis}$ between 3650 and 12500 $\rm \AA$ and  $\rm
F_{NIR}$ between 12500 and 22000 $\rm \AA$. $\rm F_{dust}$ is the total dust
flux of the galaxies.}
\begin{flushleft}
\begin{tabular}{llllll}
\hline
& & & & & \\
Name &   $\rm \log(F_{star})$ &  $\rm \log(F_{dust})$ & $\rm {F_{UV}}\over 
F_{star}$ &  $\rm {F_{vis}}\over F_{star}$ &   $\rm {F_{NIR}}\over F_{star}$ \\
& & & & & \\
\hline
NGC3034 &  -10.63  &    -10.22 &   0.03  & $~$ & $~$ \\
NGC3353 &  -12.31  &    -12.53 &   0.28  & $~$ & $~$ \\
NGC3913 &  -13.36  &    -13.74 &   0.25  &     0.56 &      0.19  \\
NGC4194 &  -12.28  &    -11.93 &   0.14  & $~$ & $~$ \\
NGC4383 &  -12.09  &    -12.28 &   0.24  & $~$ &$~$ \\
NGC4424 &  -12.10  &    -12.61 &  0.07   &    0.69  &     0.24  \\
NGC4519 &  -12.03  &    -12.54 &   0.37  &     0.49 &      0.14 \\
NGC4532 &  -12.00  &    -12.19 &   0.42  &     0.44 &      0.14 \\
NGC4670 &  -12.21  &    -12.66 &   0.50  &     0.41 &      0.09 \\          
NGC4922 &  -13.00  &    -12.48 &   0.05  &     0.65 &      0.30 \\
NGC5253 &  -11.35  &    -11.83 &   0.36  &  $~$ & $~$ \\
NGC5477 &  -12.77  &    -13.69 &   0.56  & $~$ & $~$ \\
NGC7673 &  -12.27  &    -12.52 &   0.61  & $~$ & $~$ \\          
NGC7677 &  -12.58  &    -12.61 &   0.39  & $~$ & $~$ \\          
IC732   &  -13.34  &    -12.61 &   0.03  &     0.57  &     0.40 \\           
IC3258 &  -12.48  &    -13.43 &   0.33  & $~$ & $~$ \\
IC3576 &  -12.63  &    -13.68 &   0.38  & $~$ & $~$ \\       

Z97068  &  -12.92  &    -12.83 &   0.20  &     0.55  &     0.25 \\
Z97079  &  -13.28  &    -13.57 &   0.60  &     0.32  &     0.08 \\
Z160076 &  -13.35  &    -14.15 &   0.54  &     0.40  &     0.06  \\
Z160106 &  -13.23  &    -13.40 &   0.12  &     0.62  &     0.26 \\
Z160139 &  -13.09  &    -13.62 &   0.51  &     0.45  &     0.04 \\
\hline
\end{tabular}            
\end{flushleft}
\end{table}
           
\subsection{ The comparison between the dust and stellar emission of the
starburst galaxies}

The ratio of the dust luminosity to the bolometric (stars+dust) one may  be
taken as a global measure of the extinction at least in a crude way  (e.g. 
Xu \& Buat 1995, 
Andr\'eani \& Franceschini 1996). In Fig.~5a the logarithm of this ratio is
plotted  against the absolute B magnitude of the galaxies. A clear trend is
found for brighter galaxies to be more obscured as already known (e.g. 
Giovanelli et al. 1995, Wang \& Heckman 1996). Except for 4 objects, less than
$60\%$  ($\rm \log(F_{dust}~/F_{bol}<-0.2$) of the total stellar emission is
re-emitted in the FIR range and the fraction is very low (less than 15$\%$ or
$\rm \log(F_{dust}~/F_{bol}<-0.8$) for 5 galaxies. The same variation is found 
 when the ratio of the dust luminosity to the UV one alone is considered 
 (Fig.~ 5b) as expected since this ratio is closely related to the extinction (Buat \& Xu
1996, Wang \& Heckman 1996)). 
 No trend is found with the morphological type of the galaxies.

The mean and the median of the ratio of the dust to bolometric luminosity is 
0.35 ($\sigma = 0.22$)  similar to that found for normal spiral galaxies like 
the Milky Way (around 30\%, Cox \& Mezger 1989, Xu \& Buat 1995). Therefore, in 
average a larger amount of energy is emitted by the stars than by the dust. 
 However the dispersion is very large due to two groups of extreme cases 
discussed in the previous paragraph for which the dust emission exceeds 
70$\%$ or is lower than 15 $\%$. 

The amount of energy in the UV spectral range (912-3650 $\rm \AA$) is in
average comparable to the dust emission with $\rm <F_{UV}/F_{dust} = 1.19\pm
1.38$ although the dispersion is very large; the median of $\rm 
F_{UV}/F_{dust}$ is 0.6. For the 12 galaxies for which NIR data are available 
(at least in the H band), the mean ratio of the visible (3650-12500$\rm \AA$) 
to the dust emission is $\rm <F_{vis}/F_{dust}> = 1.10\pm 0.75$ with a median 
equal to 0.9. The energy of the NIR range (12500-22000 $\rm \AA$) is only one 
third the energy emitted by the dust~: $\rm <F_{NIR}/F_{dust}> = 0.31\pm 0.20$  
with a median equal to 0.22. Once again, this low value found for the NIR 
contribution is due to the truncation between the visible and NIR ranges 
at 12500 $\AA$.

\begin{figure}
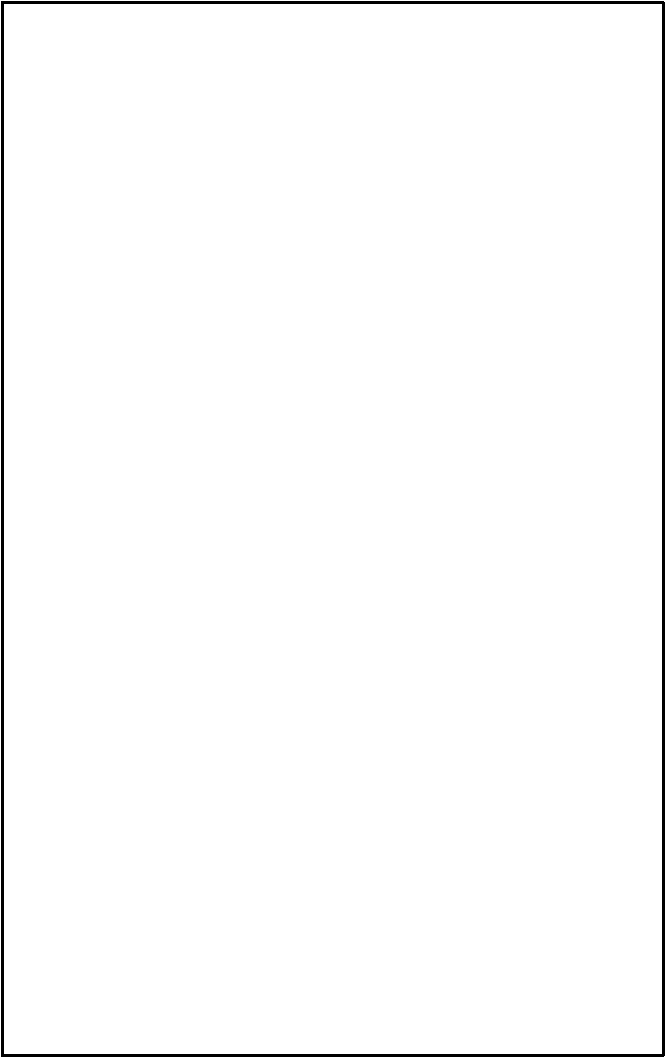

\picplace{14truecm}
\caption[]{The logarithm of the ratio of the dust to bolometric luminosities 
({\bf a}) and of the dust to UV luminosities ({\bf b}) for the 22 
starburst galaxies as a function of their absolute B magnitudes}
\end{figure}

One must be cautious in the interpretation of these results:  the spectral
energy distribution of a galaxy is governed by the star formation history and
the effects of the extinction, both beeing poorly known.  More specifically,
the old underlying population in nearby galaxies is expected   to be the major
component of the visible and NIR  emission in nearby disk  galaxies. On the 
other hand as soon as a significant star formation is  currently occurring in a
galaxy the intrinsic UV emission (i.e. without accounting for the extinction) 
is dominated by this star formation as discussed in section 4.2. 
Therefore any estimate of the extinction on the $\rm F_{dust}~/F_{bol}$ ratio 
is always very crude since the UV and visible emitting stars are not located in
the same regions of a galaxy, it is particularly true for starburst galaxies.
The use of the $\rm F_{dust}/F_{UV}$ ratio is probably more meaningfull to
estimate the extinction.  

To estimate
more quantitatively the relative contribution of the UV, visible and NIR range
to the total stellar emission, different scenarii of star formation have been
considered using the evolutionary models of Bruzual \& Charlot (1993)  (table
3). Different cases have been chosen to test the influence of an  old
population compared to that due to the stars newly formed in the starburst. 
The old disk population has been described by a constant star formation  over
10 Gyr and by an old starburst which occurred  5 Gyr
ago and lasted 1 Gyr, there are referred as CSF(10 gyr) and old SB(5 Gyr) 
 respectively in table 3.  For the current starburst we have used
two extreme cases: a very recent starburst lasting since 10 Myr (Meurer et
al. 1995) or a star formation constant over  1 Gyr (Calzetti 1997a). The models
are referred as SB(10 Myr) and CSF(1 Gyr) respectively in table 3.

\begin{table}
\caption[]{The relative constribution of the  UV, visible and NIR wavelength
range to the total star emission for different scenarii of star formation: a 
starburst (SB) lasting since 10 Myr, a constant star formation (CSF) 
over 1 Gyr and 10 Gyr and an old burst which occurred 5 Gyr ago and lasted 1
Gyr. Four combinations of these scenarii are also considered: in each case 
 the newly formed population created during the SB of 10 Myr or during the CSF
over 1 gyr accounts for 10 $\%$ of the stellar mass.  
The quantities computed are the same as in table 2}  
\begin{flushleft}
\begin{tabular}{llll}
\hline
& & &  \\
Star formation scenario &   $\rm {F_{UV}}\over 
F_{star}$ &  $\rm {F_{vis}}\over F_{star}$ &   $\rm {F_{NIR}}\over F_{star}$ \\
& & &  \\
\hline
SB (10 Myr) & 0.83 & 0.14 & 0.03\\
CSF (1 Gyr) &   0.58  & 0.34 & 0.08 \\
CSF (10 Gyr) & 0.43 & 0.43 & 0.13 \\
old SB (5 Gyr)& 0.01  & 0.69  & 0.30  \\
old SB (5 Gyr) and SB (10 Myr) & 0.82& 0.15& 0.03\\
old SB (5 Gyr) and CSF (1 Gyr) & 0.43& 0.44& 0.14\\
CSF (10 Gyr) and SB (10 Myr)& 0.82 &0.15& 0.03\\
CSF (10 Gyr) and CSF (1 Gyr) & 0.50 & 0.40 & 0.10\\ 
\hline
\end{tabular}            
\end{flushleft}
\end{table}

One can check  that the fraction of the stellar light emitted in the UV range
is large as soon as a recent star formation occurs in the galaxy.  If we
exclude the scenario of the old burst alone which cannot describe a  current
starburst,  $\rm {F_{UV}}/ F_{star}$  from the models is larger than the mean
value  $\rm <{F_{UV}}/ F_{star}>=0.32 \pm 0.19$ found for our sample (table 2)
by a factor ranging from 1.3 to 2.6.  

The contribution of the visible-NIR range becomes significant as soon as the 
galaxy   is  not experiencing a large burst with a short duration (described 
as SB(10 Myr) in  table 3 and involving 10$\%$ of the stellar mass).  Although
the statistics is poor (only 12 objects) the mean value found in table 2 for
$\rm {F_{vis}}/ F_{star}$ ($0.51 \pm 0.11$) is slightly larger than those
obtained in table 3  for the models with a current starburst described by a
constant star formation over 1 Gyr (CSF(1 Gyr)) by a factor 1.2-1.5. The
difference is much larger   when a more recent starburst is considered (SB(10
Myr)) with a factor 3.4  between the mean value of the  observations  and the
result of the models.

The comparison in the UV range  of the observations and the few models
presented here seems to favor a rather moderate extinction in our sample of 
starburst galaxies.  Under the hypothesis of  such a moderate extinction, the
analysis of the visible-NIR range argues for a starburst lasting more than 10
Myr (at the scale of the entire galaxy) to explain the rather large fraction of
visible light emerging from these galaxies.
           
\subsection{ Estimate of the extinction }

A large fraction of the light emitted by the nearby starburst galaxies studied
here is found in the UV wavelength range, at least in average.   This might
imply a moderate extinction. As discussed below, a reliable estimation  of
the extinction must account for the true spectral energy distribution of the
stellar emission as well as assuming a realistic geometry for the repartition
of dust and stars.  

To  estimate this extinction we first use a radiation transfer model in an
infinite plan parallel geometry where stars  and dust are mixed, different
scale heights are assumed for stars emitting in UV or visible. This model
treats the dust grains as frequency converters,  converting the UV and optical
radiation to the FIR one. Details on the model are given in Xu \& Buat (1995)
and  Buat \& Xu (1996).  A main result of this  model based on an energy
balance is that the UV non ionizing radiation dominates the dust heating in
disk galaxies (with starbursts or not, (Buat \& Xu 1996)).  Applying this model
to our  sample of starburst galaxies leads to  a mean extinction at 2000 $\rm
\AA$ $\rm A_{2000} =  1.33\pm 1.38$. Therefore the extinction is effectively
found to be moderate.

Actually, it has been found that the foreground screen geometry favoured by
Calzetti et al. (1994) does not fit our data probably because we deal with
integrated fluxes over the entire galaxies whereas the model of Calzetti et al.
is suited to the central regions of the galaxies where the starburst occurs
(Buat et al. 1997). Nevertheless, for the need of comparison with the high
redshift objects (see below) we also use the extinction curve of Calzetti et 
al. using an unreddened spectrum with a star formation rate constant over 1 
Gyr. Since the UV light is likely to be the major source of dust heating,
especially in starburst galaxies, we have considered the $\rm
F_{dust}~/~F_{UV}$ ratio.  Excluding the three galaxies with a $\rm
F_{dust}~/~F_{UV}$ ratio larger by a factor 30 or more than the ratios of the
other sample galaxies (see Fig. 5b), the mean value   is $\rm
<F_{dust}~/~F_{UV}>=2.5\pm 3.7$. For comparison the median value of $\rm
F_{dust}~/~F_{UV}$ for the entire sample is 1.6  The mean value of this ratio
corresponds to an  extinction  of 1.2 mag at 2000 $\rm \AA$ in fully agreement
with our previous estimate.    The extinction we find in our sample of
starburst galaxies is also consistent with the values estimated by Meurer et
al. (1995) from the UV spectral slope for a sample of 9 nearby starburst
galaxies. Such a consistency is expected  since our sample exhibits the same
range  of FIR to UV ratios as that of  Meurer et al. 

\section{Discussion and concluding remarks}

\subsection{ The local Universe}

We have considered galaxies selected to be detected photometrically both in UV
and FIR. An analysis of the local luminosity functions at both wavelengths 
shows that the galaxies selected in this way have a FIR and a UV emission
representative of the mean population of galaxies in the local Universe. From 
a sample of 22 starburst galaxies, the mean escape fraction of the stellar
light (ratio of the stellar luminosity to the bolometric one) is found to be
$63\%\pm 22\%$, i.e. similar to what is found for normal galaxies.  This escape
fraction  exhibits a strong decrease with increasing galaxy luminosity ranging
from  $\sim 80\%$ at faintest B magnitudes ($\rm M_B~>-17$) to $\sim 10\%$ for
$\rm  M_B~\sim~ -20$. This result is quite different to that of Pearson \& 
Rowan-Robinson (1996) who found that the escape fraction of the stellar light 
in starburst galaxies cannot exceed 5-10$\%$ from an analysis of deep counts 
in visible and NIR and assuming a very strong cosmological evolution for the 
starburst population. Such a low fraction of escaped light is consistent with 
the studies of FIR bright galaxies but seems not to be a generic property of 
starbursting objects, at least in the local Universe.

\subsection {Comparison with high redshift galaxies}

We can compare our results found for nearby starbursts to high redshift
galaxies recently detected. A limitation to this comparison might be that we 
deal with integrated fluxes on nearby galaxies  which contain an underlying
stellar population pre-existing to the starburst.  To our knowledge,  there is
no evidence for the   presence or not of such an older population in high
redshift starburst  galaxies (z $\sim$ 3). Nevertheless, as discussed in the paper, the UV
range is  largely dominated by the emission of the newly formed stars and the
comparison  of the high redshift  galaxies to present-day
 starburst ones  in this wavelength range  is justified. 

From an analysis of the slope of the rest-frame UV continuum, an extinction has
been estimated for  high z  galaxies (Meurer et al. 1997, Calzetti 1997b,
Pettini et al. 1997). A mean extinction of 1.65 mag at 1650 $\rm \AA$ is found
by Calzetti for a star formation rate constant over  $\sim 1$ Gyr which implies
an unreddened slope for the UV continuum $\rm \beta=-2$. This value is in
agreement with Pettini et al.'s estimates. Meurer et al. have adopted a
different star formation law with a starburst lasting 10 $\rm Myr$, leading to
a larger proportion of very young stars and a steeper UV slope ($\rm
\beta=-2.5$). Such a very extreme scenario gives an extinction as large as 3
mag. at 1600 $\rm \AA$ which can be considered as a very upper limit. The
extinction found for  nearby starburst galaxies is lower than the value
found by Calzetti or Pettini et al. at high redshift. But the galaxies 
observed at
high redshift are much more luminous that the nearby starburst galaxies studied
by us or by Meurer et al. (1995) (Meurer et al. 1997, Lowenthal et al. 1997,
Pettini et al. 1997). As an example, an extinction of 1.6 mag at 1600 $\rm \AA$
gives  $\rm F_{dust}~/~F_{bol} = 0.6$ and $\rm F_{dust}~/~F_{UV} = 4.5$ using
the extinction curve of Calzetti et al. (1994), a foreground screen and a star
formation rate constant over 1 Gyr. Extrapolating Fig.~5, these ratios correspond to the
brightest galaxies with $\rm M_B\le -20$. If the trends found in Fig.5 are
mainly due to the extinction,  the extinction estimated in high redshift star
forming galaxies is in agreement with the values found for nearby starburst
galaxies when the galaxy luminosity  is accounted for. We must point out,
however, that  low luminosity galaxies ( $\rm M_B \sim -16$) would be
detectable with a NGST-type telescope in a reasonable amount of time as far as
$\rm z\sim 9$ (if they exist). In a hierarchical scenario for the formation of
the galaxies, it is likely  that such galaxies would outnumber the large
luminous galaxies  presently detected (Ellis 1997). 

Obviously the knowledge of the local Universe properties in terms of the 
extinction occurring in galaxies as well as their spectral energy distribution, 
especially in the UV range, is essential to interpret the observations of the
Universe at high redshift. It is also crucial to predict the best way to detect
young galaxies during a  phase of intense star formation. We show that a
comparison of the FIR and UV emissions in nearby galaxies can bring some clues
since these two emissions are very sensitive to the current star formation rate
and to the extinction. A  more straightforward method would be a systematic
comparison of deep fields in UV and FIR. Such a preliminary comparison was
already performed (Buat \& Xu 1996) using FAUST (Deharveng et al. 1994) and
IRAS observations on  two 7.6$\rm ^o$-wide fields in the central region of the Virgo
cluster. Although the UV observations  of FAUST were not very deep (limiting
flux $\rm 10^{14} - 10^{15}~ erg/cm^2/s/\AA$ at 1600 $\rm \AA$), we have
searched for  galaxies detected by IRAS without any UV detected counterpart.
Given the low sensitivity of the FAUST experiment, the lower limits obtained
for  the ratio of the FIR to UV luminosity are  within the range of  values
found for the galaxies detected at both wavelengths (fig. 4  in  Buat \& Xu
1996). This is in agreement with the results found in this paper but needs to
be confirmed with   deeper observations. The UV observations carried out with
the large field  ($\rm \sim 2^o$) FOCA telescope (e.g. Donas et al. 1990) reach
the magnitude  18 at 2000$\rm \AA$. Unfortunately up to now the FIR
observations (made by the  IRAS satellite) are not deep enough to be compared
to these UV observations  and we have to wait for new FIR large field
instruments like WIRE to perform  such a comparison. 

\acknowledgements{The authors thank Alessandro Boselli and Giuseppe Gavazzi for 
giving us access to data before publication. We are also deeply grateful to
Jean Michel Deharveng for fruitful discussions and a careful reading of the
manuscript and Claus Leitherer for a helpful exchange}


\begin{thebibliography} {}
\bibitem{} 
Andr\'eani, P., Franceschini, A. 1996, MNRAS, 283, 85
\bibitem{}
Boselli, A., Tuffs, R., Gavazzi, G., Hippelein, H., Pierini, D. 1997, A\&AS 
121, 507 
\bibitem{}
Bruzual, G.A., Charlot, S. 1993, ApJ 405, 538
\bibitem{}
Buat, V. 1997, The Ultraviolet Universe at low and high
redshift, 
eds. M. Fanelli \& B. Waller, AIP conference proceedings 408, p.63
\bibitem{}
Buat, V., Burgarella, D., Xu, C. 1997, The Ultraviolet Universe at low and high
redshift,  eds. M. Fanelli \& B. Waller, AIP conference proceedings 408, p.379
\bibitem{} 
Buat, V., Xu, C. 1996, A\& A 306, 61
\bibitem{}
Buat, V., Vuillemin, A., Burgarella, D., Milliard, B., Donas,J. 1994, A\& A
281, 666 
\bibitem{}
Burigana, C., Danese, L., De Zotti, G., Franceschini, A., Mazzei, P.,
Toffolati, L. 1997 MNRAS 287, L17
\bibitem{} 
Calzetti, D. 1997a, AJ 113, 162
\bibitem{} 
Calzetti, D. 1997b, The ultraviolet Universe at low and high redshift, 
 eds. M. Fanelli \& B. Waller,  AIP conference proceedings 408, p.403
\bibitem{}
Calzetti, D., Bohlin, R., Kinney, A., Storchi-Bergman, T.,
Heckman, T. 1995, ApJ 443, 136
\bibitem{}
Calzetti, D., Kinney, A.L., Storchi-Bergmann, T. 1994, ApJ 429, 582
\bibitem{}
Cox, P., Mezger, P. 1989, A\&AR 1, 49
\bibitem{}
Deharveng, J.M., Sasseen, T.P., Buat, V., Bowyer, S., Wu ,X.  1994, A\&A 289,
715 
\bibitem{}
Devereux, N.A., Young, J.S. 1992, AJ 103, 1536
\bibitem{}
Disney, M., Davies, J., Phillips, S. 1989, MNRAS 239, 939
\bibitem{}
Donas, J., Deharveng, J.M., Milliard, B., Laget, M., Huguenin, D. 1987, A\&A
180, 12
\bibitem{}
Donas, J., Buat, V., Milliard, B., Laget, M. 1990,  A\&A 235, 60 
\bibitem{}
Donas, J., Milliard, B., Laget, M. 1995, A\&A 303, 661
\bibitem{}
Ellis, R.S. 1997, ARA\&A 35, in press
\bibitem{}
Franceschini, A., Mazzei, P., De Zotti, G., Danese, L. 1994, ApJ 427, 140 
\bibitem{}
Gavazzi, G., Boselli, A. 1996, Astrophysical letters and communications 35, 1 
\bibitem{}
Gavazzi, G., Pierini, D., Boselli, A., Tuffs, R. 1996a, A\&AS 120, 489
\bibitem{}
Gavazzi, G., Pierini, D., Baffa, C. et al. 1996b, A\&AS 120, 521
\bibitem{}
Giovanelli, R., Haynes, M., Salzer, J., Wegner, G., Da Costa, L., Freudling, W.
1995, AJ 110, 1059
\bibitem{}
Helou, G., Khan, I, Malek, L, Boehmer, L. 1988, ApJSS 68, 151
\bibitem{}
Koranyi, D.M., Strauss M.A. 1997, ApJ 477, 36
\bibitem{}
Kormendy,J., Sanders, D.B.  1992, ApJ 390, L53
\bibitem{}
Kwan, J., Xie, S. 1992, ApJ 398, 105
\bibitem{}
Lowenthal, J.D., Koo, D.C., Guzman, R., Gallego, J., Phillips, A.C., Faber,
S.M., Vogt, N.P., Garth, D.I., Gronwall, C. 1997 ApJ 481, 673
\bibitem{}
Meurer, G., Heckman, T., Leitherer, C., Kinney, A., Robert, C.,
Garnett, D. 1995, AJ 110, 2665
\bibitem{}
Meurer, G., Heckman, T., Lehnert, M., Leitherer, C., Lowenthal, J. 1997, AJ
114, 54 
\bibitem{}
Mazzei P., De Zotti, G., Xu, C.  1994,  ApJ 422, 81
\bibitem{}
Milliard, B., Viton, M., Martin, C., Donas, J. 1997, The Ultraviolet Universe 
at low and high redshift, eds. M. Fanelli \& B. Waller, AIP conference
proceedings 408, p.106
\bibitem{}
Mobasher, B., Rowan-Robinson, M., Georgakakis, A., Eaton, N. 1996, MNRAS 282,
L7 
\bibitem{}
Pearson, C., Rowan-Robinson, M. 1996 MNRAS 283, 174
\bibitem{}
Pettini, M., Steidel, C., Dickinson, M., Kellog, M., Giavalisco, M.,
Adelberger, K. 1997, The ultraviolet Universe at low and high redshift, 
 eds. M. Fanelli \& B. Waller, AIP conference
proceedings 408, p.279
\bibitem{}
Sanders, D.B., Mirabel, I.F. 1996, ARA\&A  34, 749
\bibitem{}
Satyapal, S., Watson, D., Pipher, J., Forrest, W., Greenhouse, M., Smith, H.,
Fisher, J., Woodward, C. 1997, ApJ 483, 148
\bibitem{}
Saunders, W., Rowan-Robinson, M., lawrence, A., Efstathiou, G., kaiser, N.,
Ellis, R.S., frank, C.S. 1990, MNRAS 242, 318
\bibitem{}
Smith ,D., Neff, S., Bothun, G. et al. 1996, ApJ 473, L21
\bibitem{}
Soifer B.T., Neugebauer T. 1991, AJ 101, 354
\bibitem{}
Steidel, C.C., Giavalisco,G.,Dickinson, M., Adelberger, K. L. 1996a AJ 112, 352
\bibitem{}
Steidel, C.C., Giavalisco,G., Pettini, M., Dickinson, M., Adelberger, K. 1996b 
ApJ 462, L17 
\bibitem{}
Trewhella, M., Davies, J., Disney, M., Jones, H. 1997, MNRAS 288, 397
\bibitem{} 
Treyer, M.A., Ellis, R.S., Milliard, B., Donas, J. 1997, The Ultraviolet 
 Universe at low and high redshift, eds. M. Fanelli \& B. Waller, p.99
\bibitem{}
Tully, R.B. 1988, Nearby Galaxies Catalog,  Cambridge University Press
\bibitem{}
van der Werf, P.P., Israel, F.P. 1996, Science with large millimeter arrays, 
ed. P. Shaver, Springer
\bibitem{}
Vaucouleurs de, G., Vaucouleurs de, A., Corwin, H.G., Buta, R.J., Paturel, G., 
Fouqu\'e , P. 1991, Third Reference Catalogue og Bright Galaxies (New York: 
Springer Verlag)
\bibitem{}
Wang, B., Heckman, T. 1996, ApJ 457, 645
\bibitem{}
Wright, G, Joseph, R., Robertson, N., James, P., Meikle, W. 1988, MNRAS 233, 1
\bibitem{} 
Xu, C., Buat, V. 1995 A\&A 293, L65

\end{thebibliography}
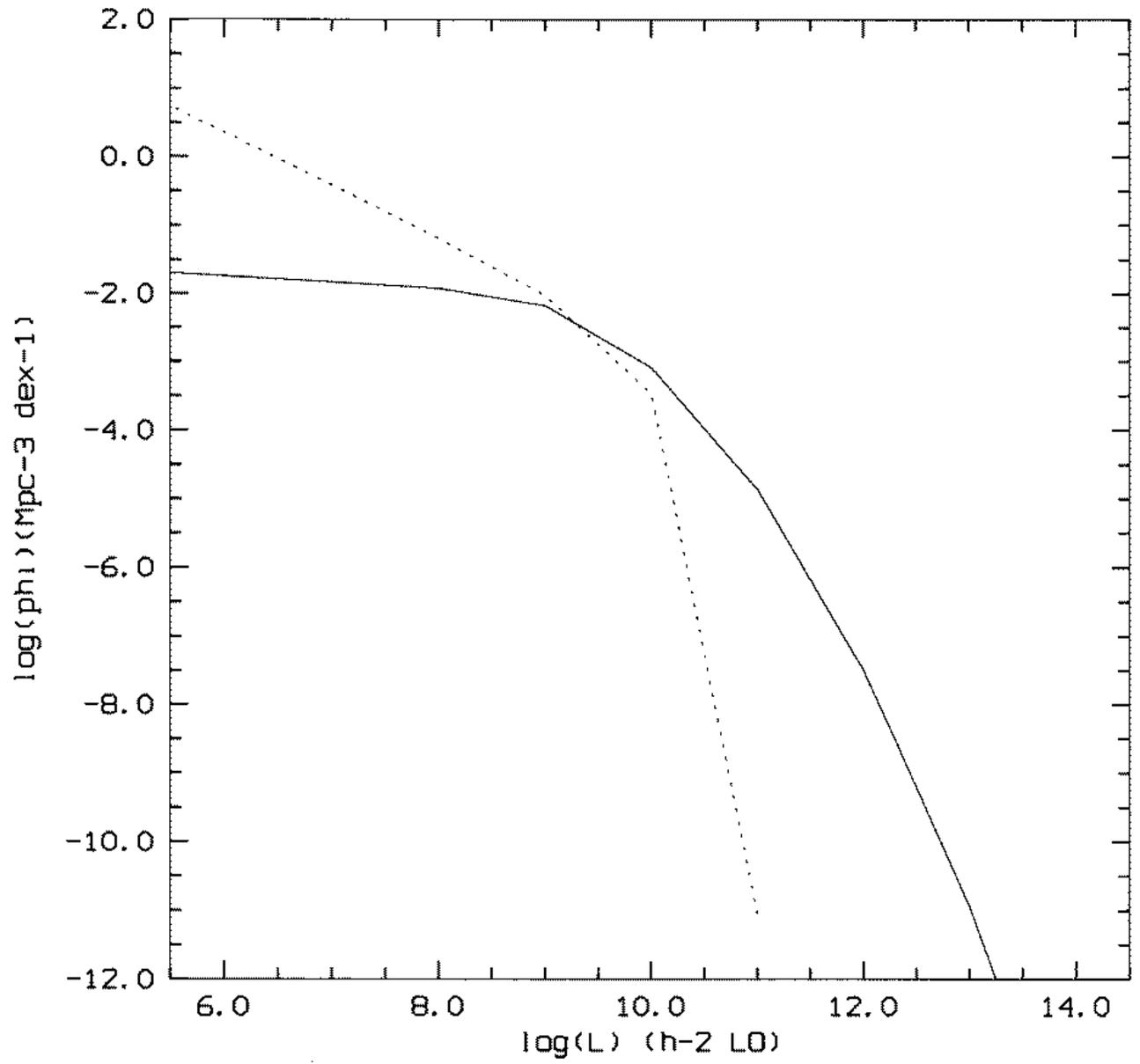
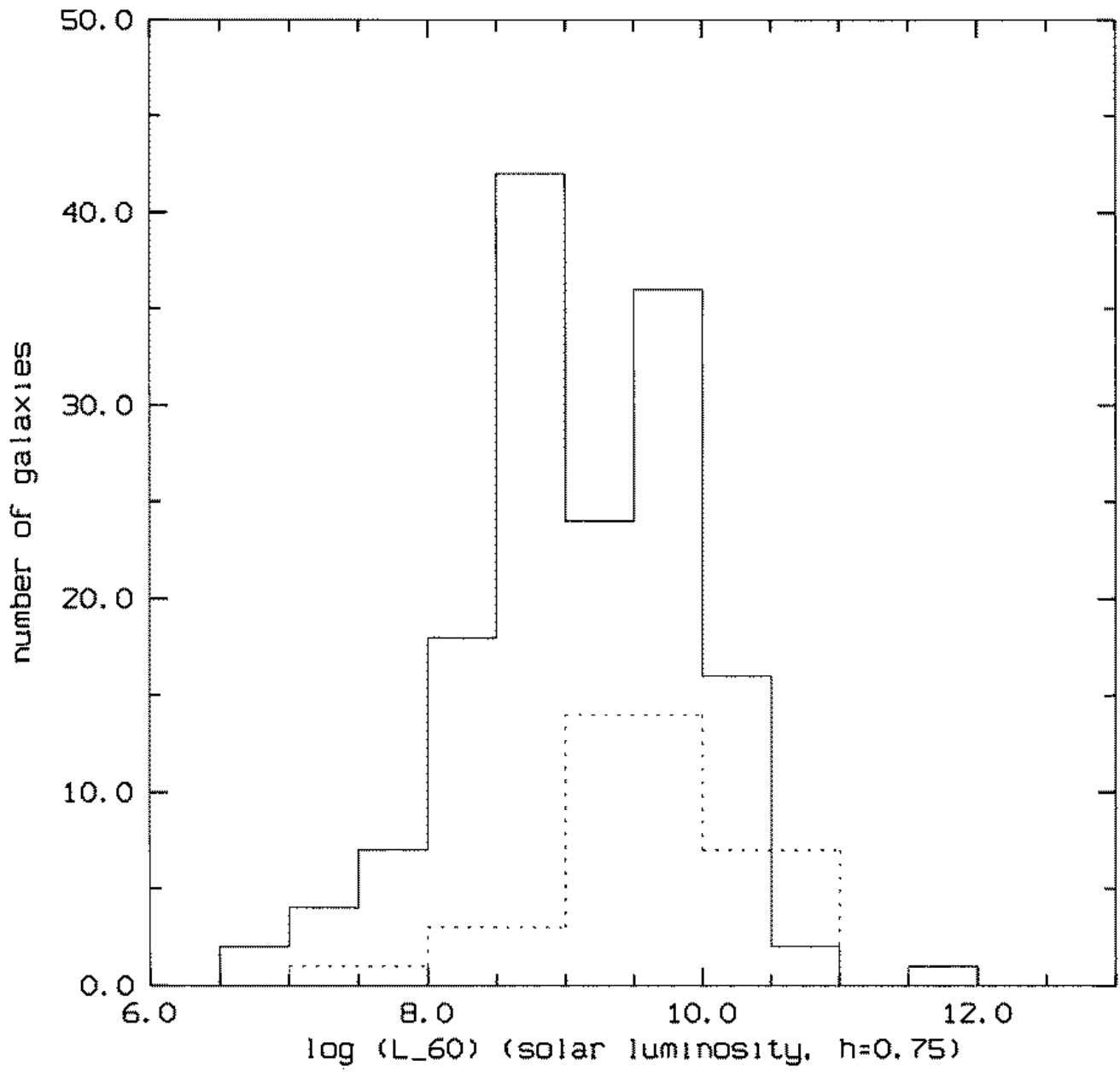
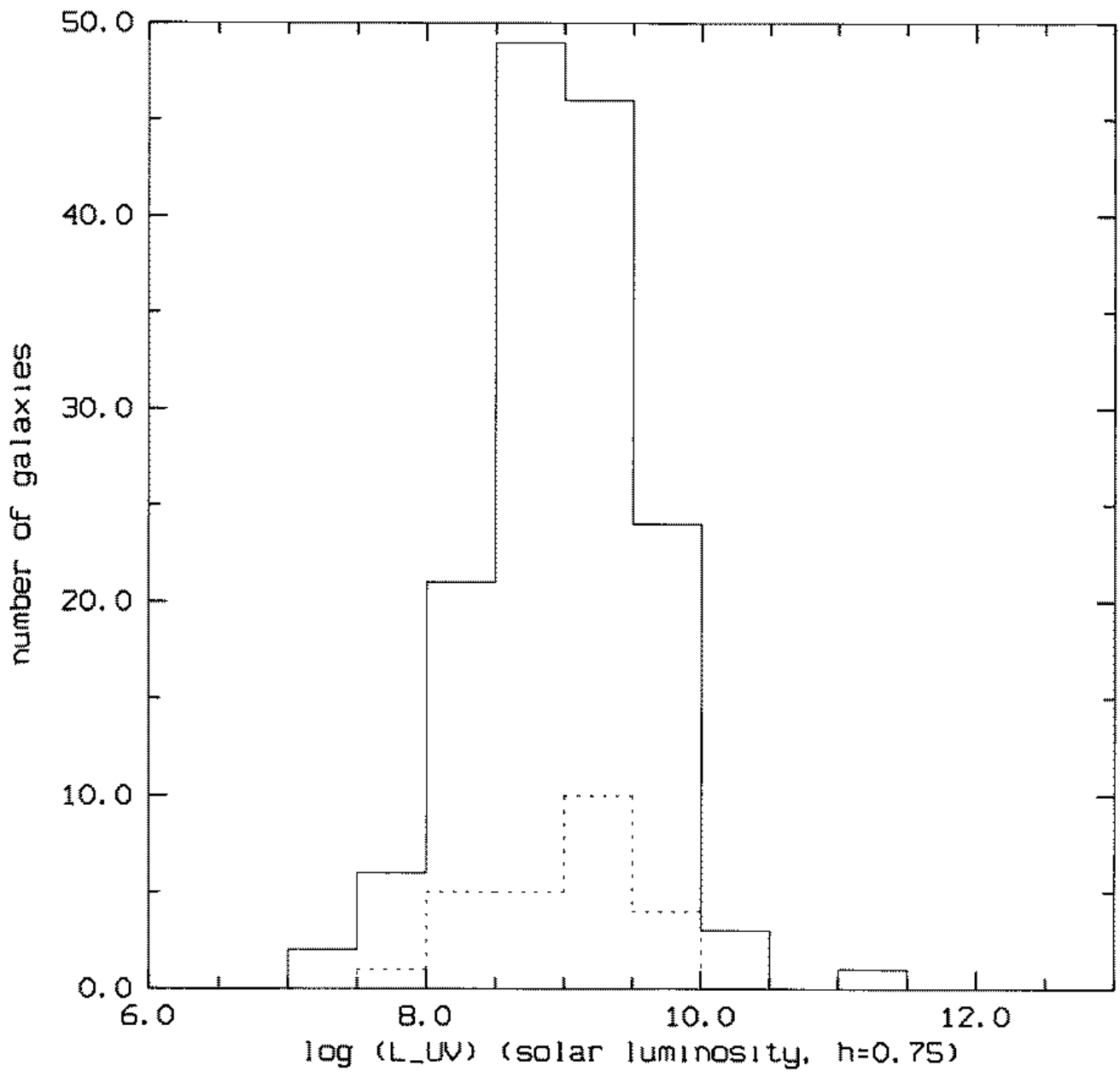
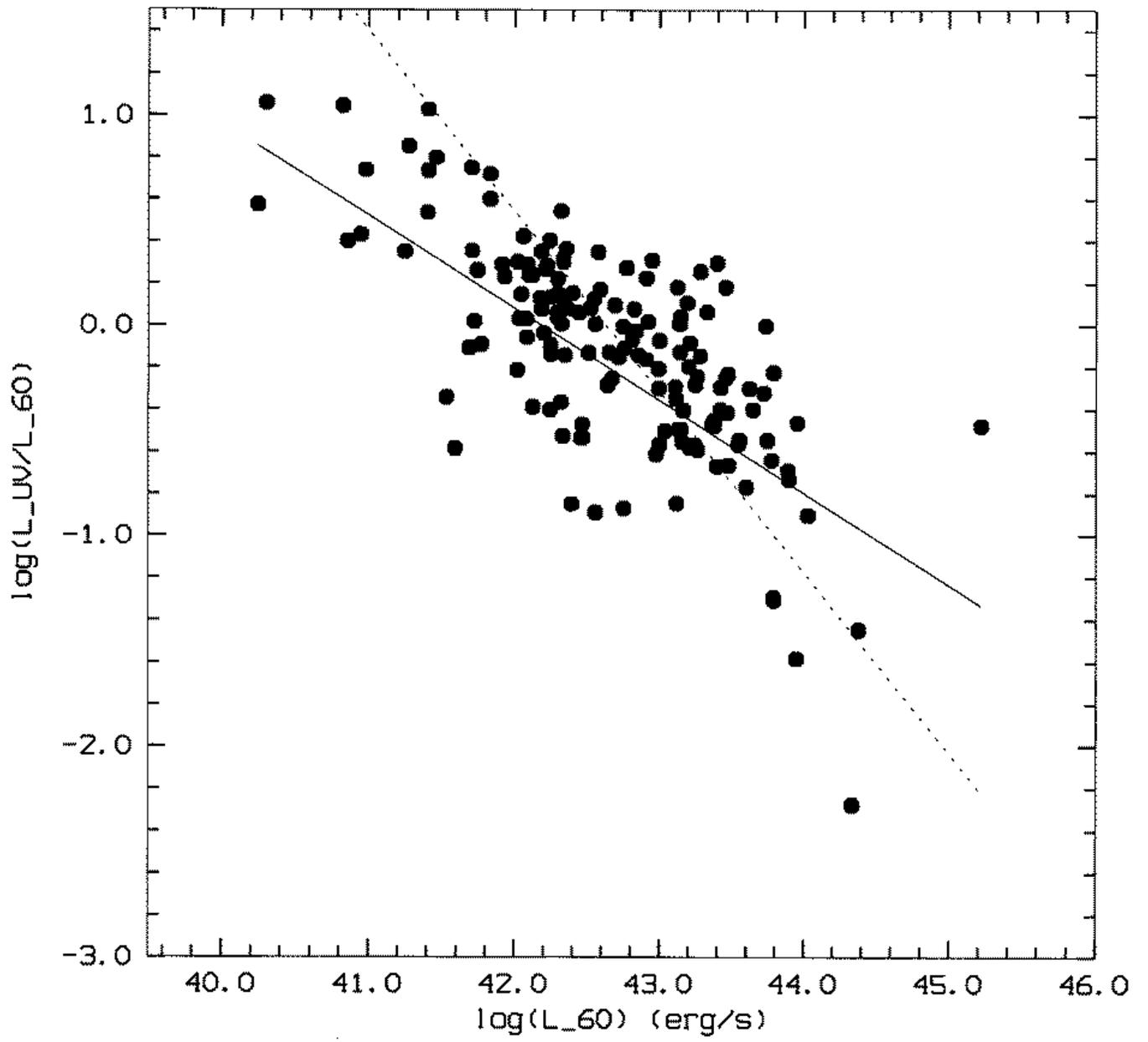
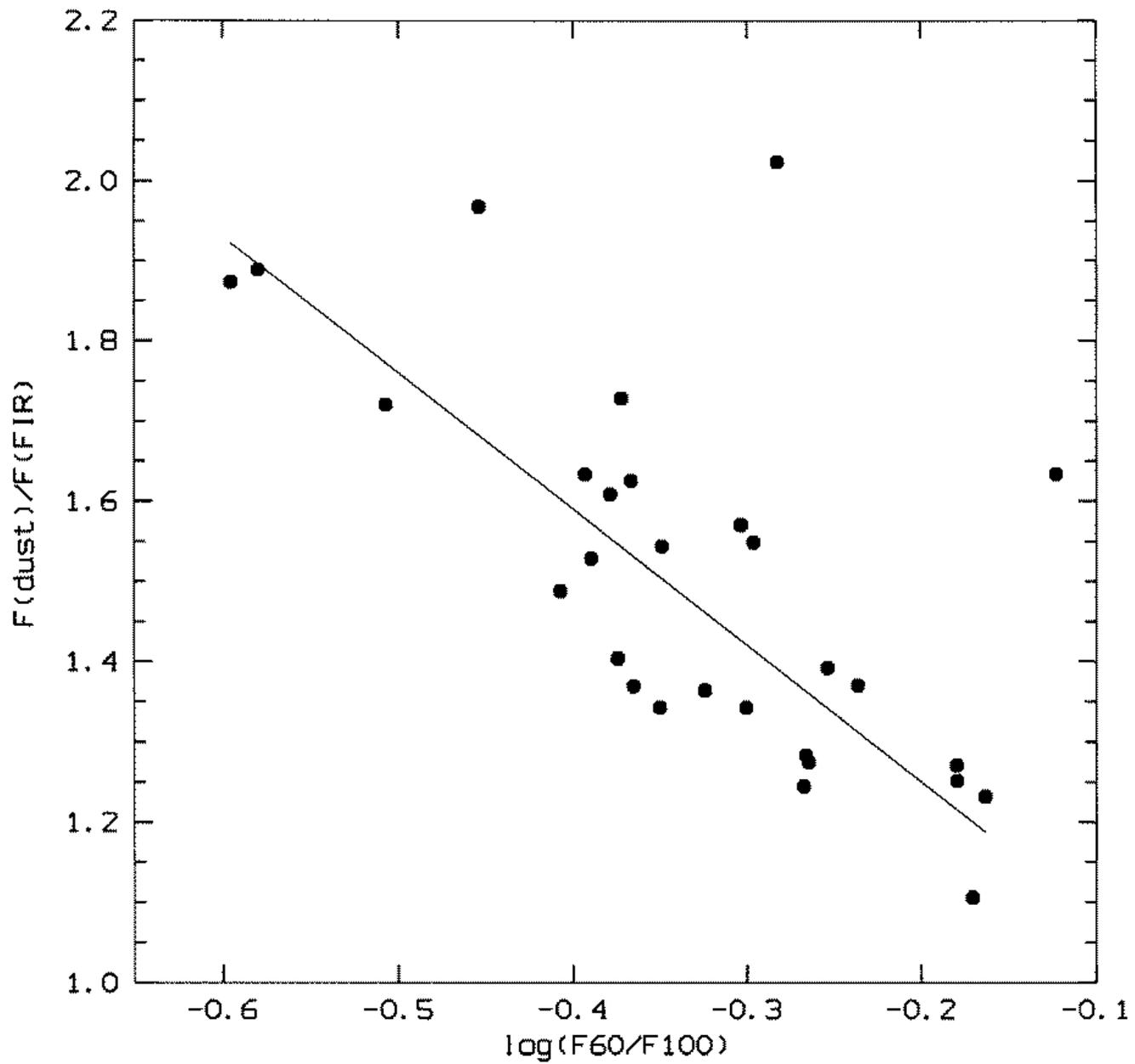
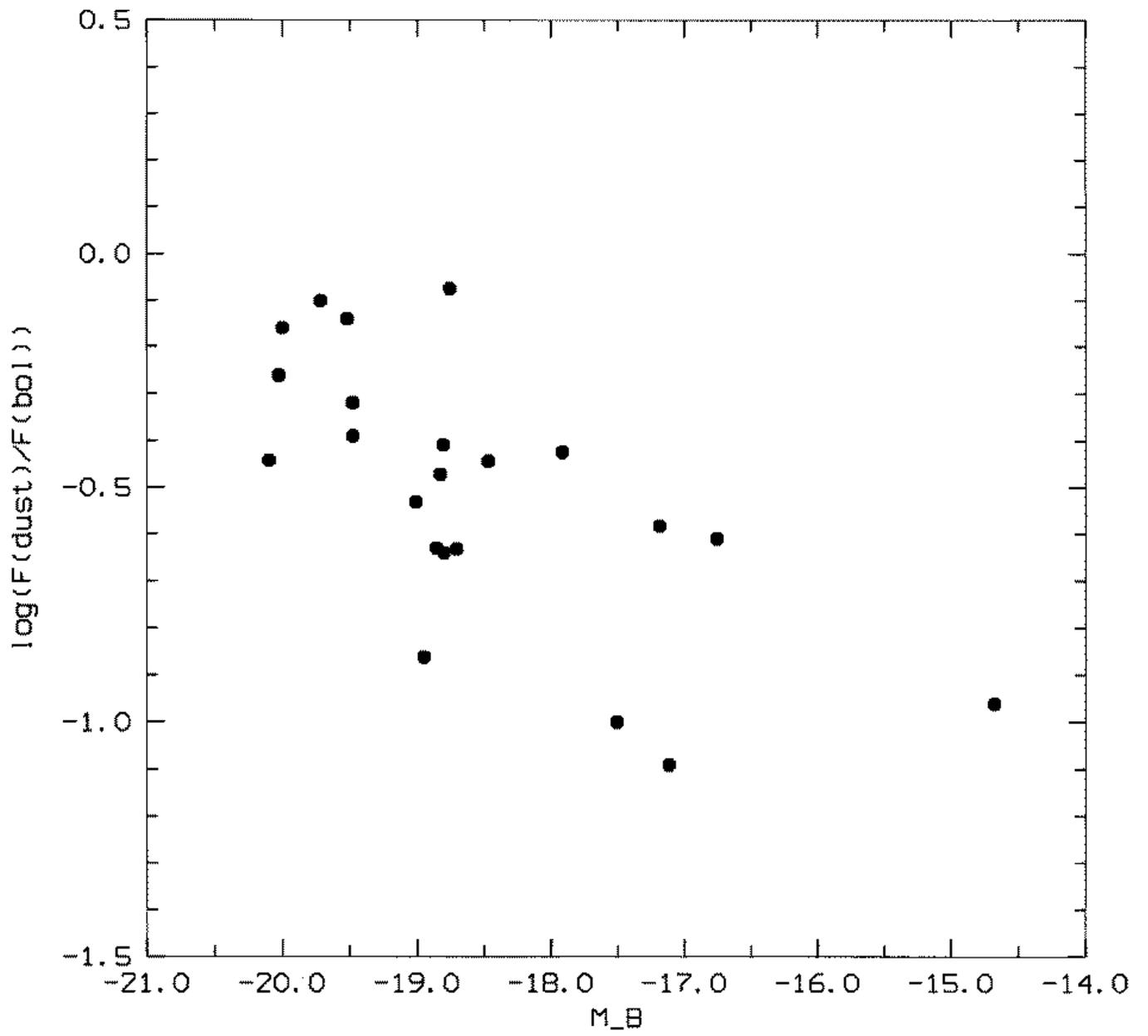
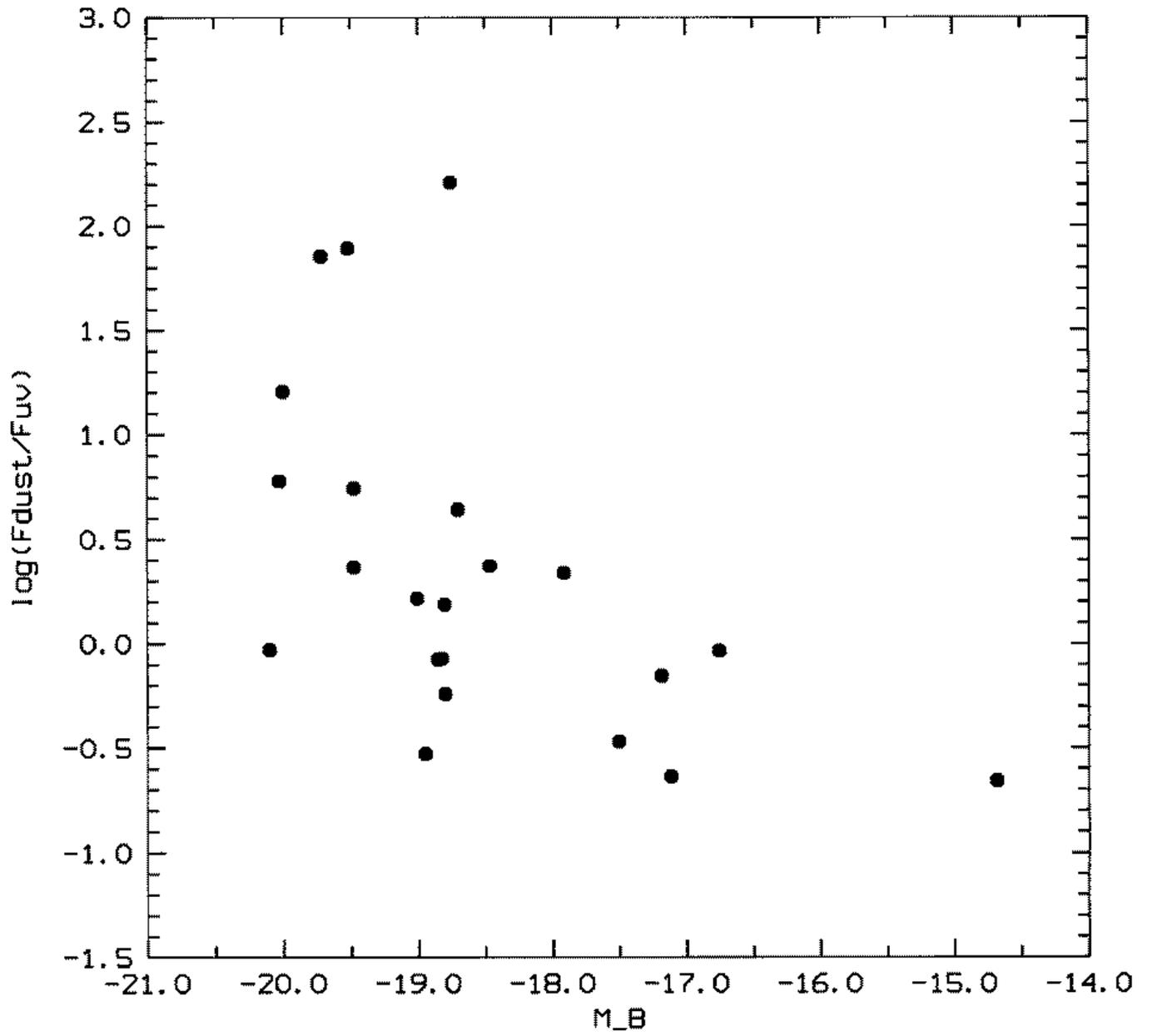
\end{document}